\documentclass[preprint2]{aastex6}
\usepackage{epsfig}
\usepackage{amsfonts,amssymb,amsmath}

\begin{document}

\title{Theoretical Predictions of Colors and Metallicity of the Intra-Cluster Light}
\author{E. Contini$^{1}$, S.K. Yi$^{1},$, X. Kang$^{2}$}

\affil{$^1$Department of Astronomy and Yonsei University Observatory, Yonsei University, Yonsei-ro 50, Seoul 03722, Republic of Korea}
\affil{$^2$Purple Mountain Observatory, the Partner Group of MPI f\"ur Astronomie, 2 West Beijing Road, Nanjing 210008, China}

\email{contini@pmo.ac.cn}
\email{yi@yonsei.ac.kr}
\email{kangxi@pmo.ac.cn}

\begin{abstract} 

We study colors and metallicities of the Brightest Cluster Galaxies (BCGs) and Intra-Cluster Light (ICL) in galaxy groups and clusters, as predicted by a semi-analytic model of galaxy 
formation, coupled with a set of high-resolution N-body simulations. The model assumes stellar stripping and violent relaxation processes during galaxy mergers to be the main channels 
for the formation of the ICL. We find that BCGs are more metal-rich and redder than the ICL, at all redshifts since the ICL starts to form ($z\sim 1$). In good agreement with several 
observed data, our model predicts negative radial metallicity and color gradients in the BCG+ICL system. By comparing the typical colors of the ICL with those of satellite galaxies,
we find that mass and metals in the ICL come from galaxies of different mass, depending on the redshift. Stripping of low mass galaxies, $9<\log M_* <10$, is the most important contributor
in the early stage of the ICL formation, but the bulk of the mass/metals contents are given by intermediate/massive galaxies, $10<\log M_* <11$, at lower redshift. 
Our analysis supports the idea that stellar stripping is more important than galaxy mergers in building-up the ICL, and highlights the importance of colors/metallicity measurements for 
understanding the formation and evolution of the ICL.

\end{abstract}


\section[]{Introduction} 
\label{sec:intro}

The intracluster light (ICL), which was first predicted by \cite{zwicky37} and the observed in the Coma Cluster by \cite{zwicky51}, constitutes an important component of baryonic matter in galaxy groups and clusters. Given its intrinsic nature, i.e. 
diffuse light made up of stars not bound to any galaxy, understanding its formation and evolution is fundamental for understanding the dynamical history of the group/cluster in which it resides. 
Most of the ICL is concentrated around the brightest cluster galaxy (BCG), but a non-negligible fraction is found around intermediate/massive satellites (\citealt{presotto14,contini14,contini18}), especially 
in massive ($\sim 10^{15} M_{\odot}$) clusters at redshift $z=0$. Despite the physical processes at play for the formation of this diffuse light are still under debate, there is a general consensus 
that ICL and BCGs are linked in their formation and evolution (\citealt{murante07,purcell07,puchwein10,rudick11,contini14,demaio15,burke15,groenewald17,morishita17,montes18,demaio18,contini18}), in 
particular after $z \sim 0.7$, during which a clear co-evolution between the two components has been found in our recent study (\citealt{contini18}, hereafter C18).

In C18 we discussed the relative importance of the two main processes that are believed to be responsible for the formation of the ICL, i.e. stellar stripping and galaxy mergers. These two processes 
have different consequences on the ICL properties, such as colors and metallicities. Qualitatively speaking, if mergers are mainly responsible for the formation of the ICL, we would expect no color/metallicity 
gradients in the BCG+ICL system, which translates in similar colors/metallicities of BCGs and ICL (as argued in C18, but see also \citealt{montes14,demaio15,morishita17,montes18}). On the 
other hand, if stellar stripping of satellite galaxies is the main contributor to the ICL, it is reasonable to expect some kind of radial gradients of BCG+ICL colors and metallicity 
(\citealt{demaio15,demaio18}). The most recent observations (\citealt{morishita17,iodice17,demaio18,montes18}, just to quote a few of them) are finding clear radial gradients for both colors and metallicity,
in a wide range of redshift, a clue which favors stellar stripping rather than galaxy mergers as the dominant mechanism for the ICL formation. 

In \cite{contini14} (hereafter C14) we argued that measurements of the ICL metallicity can help to constrain theoretical models. During the last few years the observational measurements of the ICL (or BCG+ICL) 
remarkably increased, in such a way that it is now possible to test model predictions. \cite{montes14} derive the stellar population properties of the ICL in Abell Cluster 2744, a massive 
cluster at $z\sim 0.3$. From the restframe colors of the ICL, they derive a mean metallicity comparable with the solar value, and a metallicity gradient of the global BCG+ICL system. Similarly, \cite{demaio15}
analyse four galaxy clusters in the redshift range $0.44<z<0.57$, and for three of them they find a clear metallicity gradient from super-solar metallicities in the region dominated by the BCG, to sub-solar 
metallicites in the region dominated by the ICL. A few years later, these results have been confirmed by other studies (e.g., \citealt{demaio18,montes18}), and so strengthening the idea that 
stellar stripping could be the dominant contributor of the ICL. 

Observations focused also on the BCG+ICL colors. \cite{demaio18} extend the analysis done in \cite{demaio15} by studying the ICL properties of 23 galaxy groups and clusters in the redshift range $0.29<z<0.89$,
and focusing mainly on colors. They find that the color gradients of the BCG+ICL systems become bluer with increasing radial distance, and argue that this cannot be the result of violent relaxation processes 
during major mergers between satellite galaxies and the BCG. Moreover, they conclude that tidal stripping of massive galaxies ($\log M_* /M_{\odot} > 10.4$) in the very vicinity of the group/cluster centre 
($<100$kpc) is the likely source of the ICL, in good agreement with our results in C14 and C18. Similar results have been found in \cite{morishita17}. These authors investigate the ICL properties of six clusters 
in the redshift range $0.3<z<0.6$ and find negative color gradients with increasing radial distance from the BCG. However, from the typical colors of the satellite population, they conclude that the ICL 
likely originated from satellites with mass $\log M_* /M_{\odot} < 10$, in contrast with our results (C14) and \cite{demaio18}. Their conclusion is in contrast also with \cite{montes18},
who use the Hubble Frontier Fields survey to analyse the properties of the ICL in six massive clusters at redshift $0.3<z<0.6$. They find that the average ICL metallicity 
([Fe/H] $\sim -0.5$) is compatible with that of the outskirts of the Milky Way, and the mean stellar ages of the ICL are younger (between 2-6 Gyr) than the most massive galaxies in the clusters, suggesting that the ICL form mainly from stripping of intermediate (Milky Way like) galaxies after $z<1$.

Very recently, \cite{ko18} analyse the amount of ICL and its properties in a 
cluster at $z\sim 1.3$ (which is the highest redshift for which spatial distribution, colors and quantity of ICL are available, as yet) and they report no radial dependence of the ICL color.

In this paper we take advantage of the model for the ICL formation described in C14 and C18, and a variation of it, to fully analyse the ICL and BCG colors and metallicity. In the original version of the model,
during an episode of stellar stripping we assume that the same fraction of stellar mass and metals is moved to the ICL component, i.e. we assume no metallicity gradient in satellite galaxies. A modified version 
of the model presented here assumes a random (negative) metallicity gradient in satellites. Our analysis will focus on addressing the following points:
\begin{itemize}
 \item [i)] to make model predictions of colors and metallicites and quantify the difference between the two models;
 \item [ii)] to compare our predictions with the available observed data;
 \item [iii)] to study the redshift evolution of ICL colors and metallicity and use the color-color and metallicity-color planes to analyse the contribution to the ICL from satellite of different mass as 
 a function of redshift.
\end{itemize}

In Section \ref{sec:methods}, we briefly summarise our model for the formation of the ICL presented in C14 and C18 and its modification. In Section \ref{sec:results}, we show the results of our analysis, which 
will be discussed in detail in Section \ref{sec:discussion}. In Section \ref{sec:conclusions}, we give our conclusions. Throughout this paper we adopt a standard $\Lambda$CDM cosmology assuming the following cosmological 
parameters: $\Omega_m=0.24$ for the matter density parameter, $\Omega_{\rm bar}=0.04$ for the contribution of baryons, $H_0=72\,{\rm km\,s^{-1}Mpc^{-1}}$ for the present-day Hubble constant, $n_s=0.96$ for the primordial 
spectral index, and $\sigma_8=0.8$ for the normalization of the power spectrum. Stellar masses (with the assumption of \citealt{chabrier03} IMF) are given in units of $M_{\odot}$ (unless otherwise stated), while magnitudes 
are in the AB system.

\section[]{Methods}  
\label{sec:methods}

In this section we briefly summarise our modelling for the ICL formation and describe its modification, which have been implemented in the semi-analytic model presented in \cite{delucia07}. As in C14 and C18, the 
semi-analytic model has been coupled with the same set of high-resolution N-body simulations (\citealt{contini12}). For further information about the semi-analytic model and the details of the set of simulations, we refer 
the reader to \cite{delucia07}, C14 and C18.

We take advantage of two models: for the sake of simplicity we name them {\small STANDARD} and {\small METGRAD}. The {\small STANDARD} model is formally identical to the Tidal Radius + Merg. model adopted in C14 
and the ``STANDARD'' model adopted in C18. For those readers not familiar with the model, we provide here the necessary information for a full understanding.

This model takes into account the tidal forces between satellite galaxies and the potential well of the group/cluster within which they reside, and violent relaxation during galaxy mergers. The tidal forces are 
responsible for the stellar stripping, which means that part of the stellar mass (sometimes all) of the satellite galaxy that suffers the tidal force is stripped from the galaxy and gets unbound. This mass is assumed 
to move to the ICL component associated to the central galaxy at the moment of the stripping. Clearly, this model of stellar stripping allows satellite galaxies to lose mass in a continuous fashion, before merging or 
being disrupted if the tidal field is strong enough. The stellar density profile of the simulated satellites is approximated by a spherically symmetric isothermal profile, such that the \emph{tidal radius} can be estimated 
by means of the equation (see \citealt{binney08}):
\begin{equation}\label{eqn:tid_rad}
  R_{t} = \left(\frac{M_{sat}}{3 \cdot M_{DM,halo}}\right)^{1/3} \cdot D \, ,
\end{equation} 
where $M_{sat}$ is the satellite mass (stellar mass + cold gas mass), $M_{DM,halo}$ is the dark matter mass of the parent halo, and $D$ the satellite distance from the halo centre. 

An isothermal profile is assumed by the semi-analytic model to derive the tidal radius via Equation \ref{eqn:tid_rad}. However, for a more realistic implementation of stellar stripping, a satellite galaxy is considered 
to be a two-component system with a spheroidal component (the bulge), and a disk component, when stellar stripping occours. If $R_t$ is smaller than the bulge radius, the satellite is assumed to be completely destroyed 
and its stellar and cold gas mass to be added to the ICL and hot component of the central galaxy, respectively. On the other hand, if $R_t$ is larger than the bulge radius but smaller than the disk radius, only the stellar 
mass in the shell $R_t -R_{sat}$ is moved to the ICL component (as well as a proportional fraction of the cold gas to the hot component of the central galaxy). Since we model the disk component with an exponential profile,
$R_{sat} = 10 \cdot R_{sl}$, where $R_{sl}$ is the disk scale length, is the radius which contains 99.99 per cent of the stellar mass in the disk. After an episode of partial stripping (no disruption), we update $R_{sl}$ 
to one tenth of $R_t$.

It must be noted that our semi-analytic model distinguishes two kinds of satellites, type1 and type2 (a.k.a. orphans) satellites. The difference between the two types relies on the dark matter content: type1 
galaxies still own their parent subhalo, while type2 have lost their parent subhalo or it went under the resolution of the simulation. For type2 the semi-analytic model applies Eq. \ref{eqn:tid_rad} above directly with no 
other filter, but for type1 satellites, the model first requires that the following condition is met:

\begin{equation}\label{eqn:eq_radii}
 R^{DM}_{half} < R^{Disk}_{half} \, ,
\end{equation}
where $R^{DM}_{half}$ is the half-mass radius of the parent subhalo, and $R^{Disk}_{half}$ the half-mass radius of the galaxy's disk, that is $1.68\cdot R_{sl}$ for an exponential profile. 
Central galaxies can also increase their associated ICL through accretion of ICL originally associated to satellite galaxies. This mechanism of ICL accretion works 
differently depending on the type of satellite involved. Centrals accrete ICL originally associated to satellite galaxies that pass from type1 to type2 \footnote{Orphans are  
not allowed to carry any ICL in our model.}. Moreover, anytime a type1 satellite is affected by stellar stripping, its associated ICL is added to the ICL component of the corresponding central galaxy. In order to summarise, a central galaxy acquires its ICL due to stellar stripping through three 
mechanisms:
\begin{itemize}
 \item direct stripping of the stellar mass of satellite galaxies;
 \item accretion of ICL associated to type1 satellites which experience a stripping event;
 \item accretion of ICL associated to satellite galaxies which pass from type1 to type2 (i.e. they have lost their dark matter component).
\end{itemize}

Tidal stripping is not the only channel from which the ICL can form in the {\small STANDARD} model. Similarly to C14 and C18, we consider also violent relaxation processes that take place during mergers. The ``merger channel" is modelled as follows: at each merger, we assume that a fraction $f_m =0.2$ of the satellite stellar mass becomes unbound and is added to the ICL of the corresponding central galaxy. The fraction $f_m$ has been set to 0.2 by means of  numerical simulations of groups (\citealt{villalobos12}). We have verified that such a simple prescription reproduced the result of the simulations, although in the reality $f_m$ is expected to depend on the circularity of the orbit, or other 
satellite properties such as its stellar mass.

In this study we also consider a slight modification of the {\small STANDARD} model, that we named {\small METGRAD} model. The {\small STANDARD} model does not assume any metallicity gradient in satellite galaxies 
when they are subject to stellar stripping, which means that the same fraction of stellar mass and metals are stripped from the galaxy. The novelty in the {\small METGRAD} model relies on the assumption of a (negative) 
metallicity gradient in satellites \footnote{Central galaxies are not subject to stellar stripping, i.e. no assumption is maden for their metallicity profile.} (see, e.g. \citealt{tissera17}), such that the fraction of metals stripped is different from the fraction of stellar mass stripped. In this model, at each episode of stripping we 
assume that metals in the satellite galaxies follow an exponential profile with $R_{sl,metals}=f_R \cdot R_{sl}$, where $R_{sl,metals}$ is the scale lenght of the distribution of metals, and $f_R$ is a random fraction 
assumed to be between 0.5 and 1. Then, in the {\small METGRAD} model metals are on average more concentrated in the inner regions of the disk, with respect to the {\small STANDARD} model. As a natural consequence, the ICL 
from the {\small METGRAD} model is expected to be less metal-rich than the ICL in the {\small STANDARD} model.

\section{Results}
\label{sec:results}

In this section we present the predictions of our models for the ICL and BCG colors and metallicities. We will focus on these two properties, their variation as a function of redshift, and the differences between 
the two models that will be tested against available observational data.

\subsection{ICL and BCG metallicities}
\label{sec:metallicity}

\begin{figure*} 
\begin{center}
\begin{tabular}{cc}
\includegraphics[scale=.45]{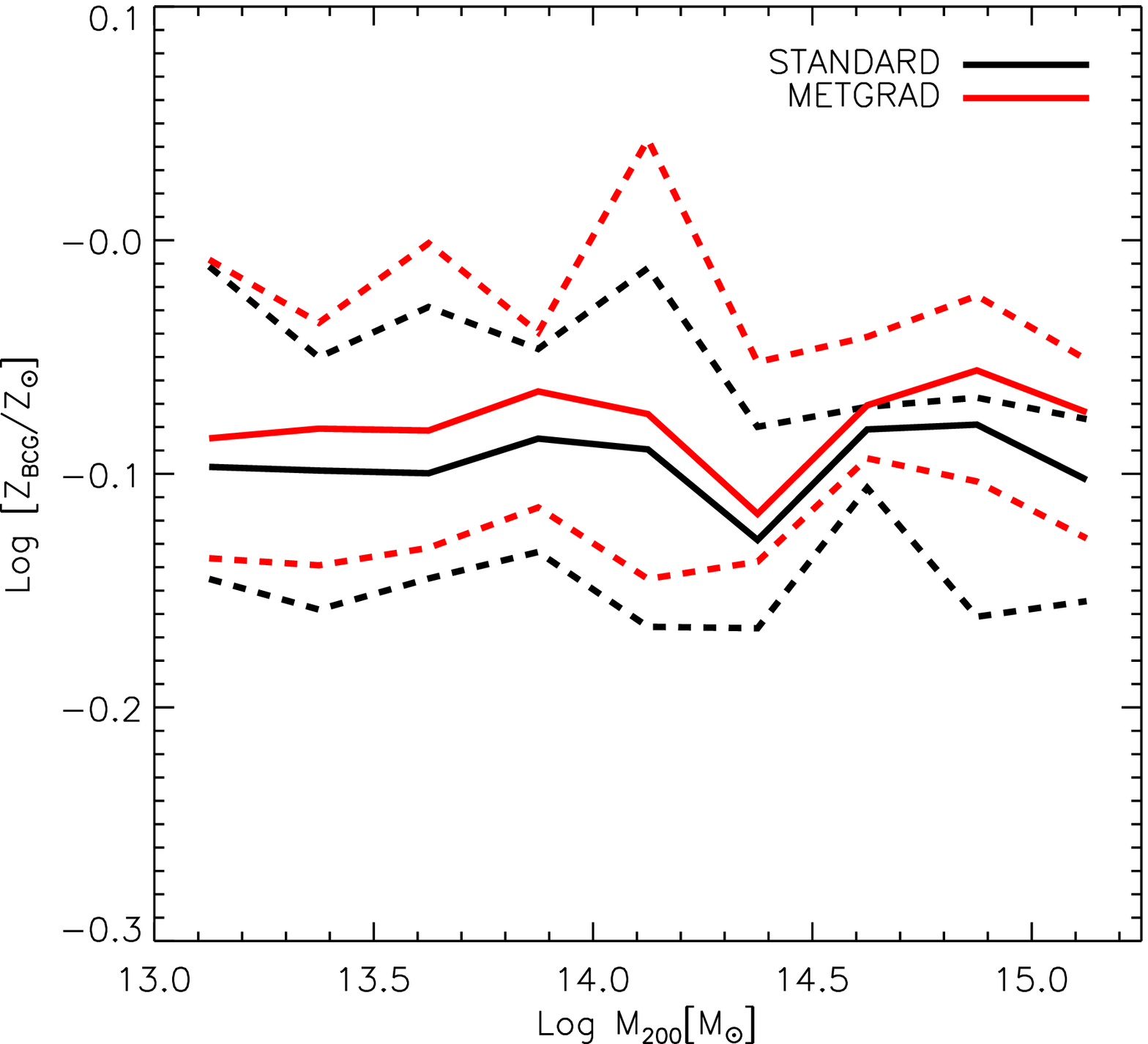} &
\includegraphics[scale=.45]{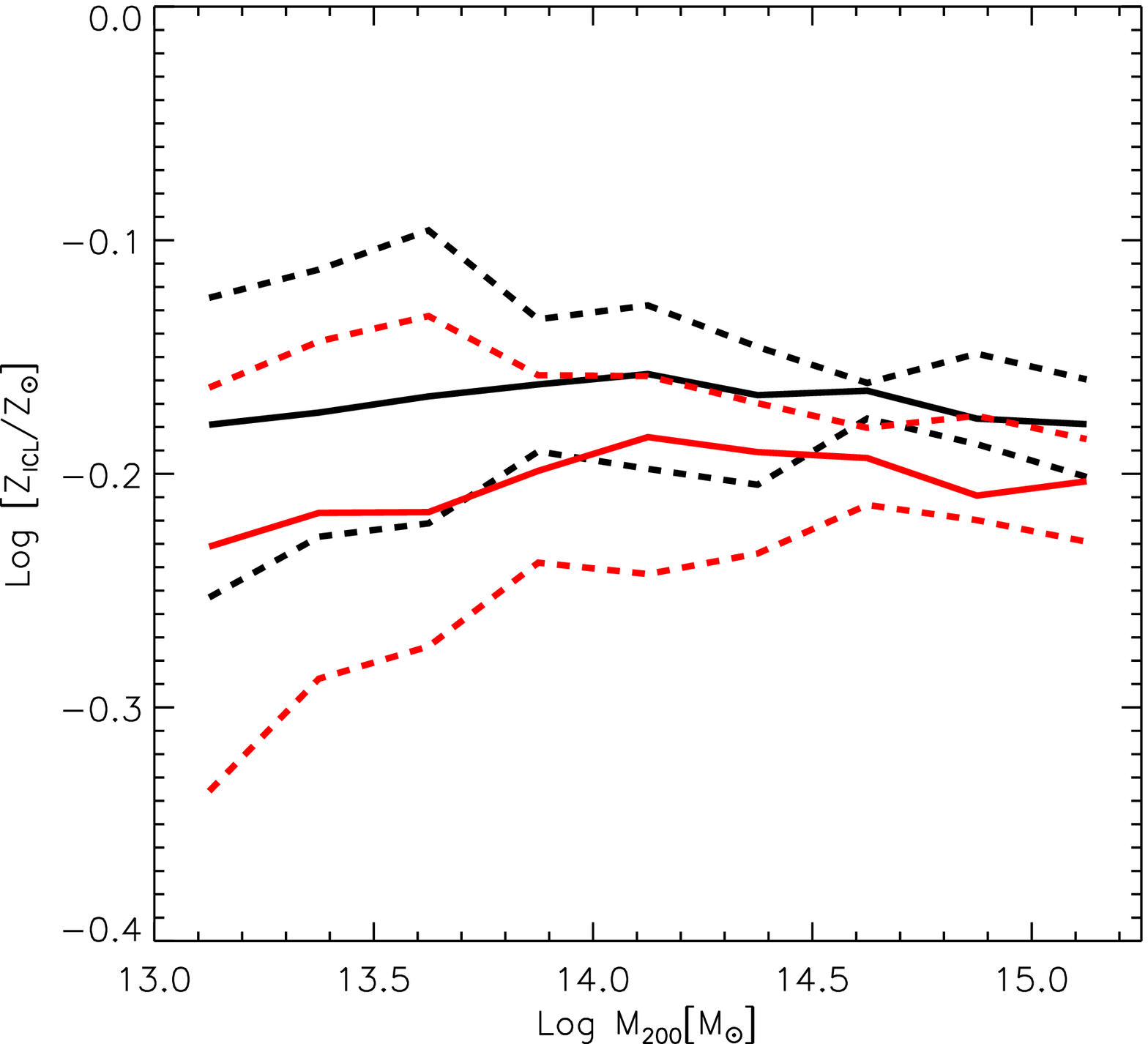} \\ 
\end{tabular}
\caption{Metallicity of BCGs (left panel) and ICL (right panel) as a function of halo mass, as predicted by our models (color lines as indicated in the legend). Solid lines represent the median of the distribution, 
and dashed lines represent the 16$^{th}$ and 84$^{th}$ percentiles.}
\label{fig:bcgicl_met}
\end{center}
\end{figure*}

In C14 we showed that the metallicity of the ICL and BCGs do not depend on the halo mass (see Fig.12 and Fig.13 of C14), and argued that detailed observational data of these 
properties could help in constraining models. After a few years, a non-negligible amount of data has been collected and it is possible to test our model predictions against them. 

However, before testing our predictions with observed data, in Figure \ref{fig:bcgicl_met} we show the metallicity of BCGs (left panel) and the metallicity of the ICL (right panel) as a function of halo mass, 
as predicted by the {\small STANDARD} (black solid lines) and {\small METGRAD} (red solid lines) models. The dashed lines represent the 16$^{th}$ and 84$^{th}$ percentiles of the distributions. As expected and 
anticipated in Section \ref{sec:methods}, model {\small METGRAD} predicts slightly higher metallicities for the BCGs, and lower metallicities for the ICL. This is a consequence of the fact that, in model 
{\small METGRAD}, metals in satellite galaxies are more concentrated in the inner part of the disk. In this model, the metallicity of the ICL decreases because the stellar mass moved after an episode of stripping 
is less metal-rich (with respect to the case of the {\small STANDARD} model). As time passes and the ICL grows and evolves, less metals are deposited in the ICL. On the other hand, satellites which suffer from stellar 
stripping but not destroyed, will survive being more metal-rich. As they merge with the BCG, they bring a higher amount of metals, thus increasing the metallicity of the BCG. It must be noted, however, that the 
difference between the two models is basically negligible in both cases, BCGs and ICL. In fact, the BCG metallicity on average increases of only $\sim 0.02$ dex, while the net average decrease in the ICL is 
$\sim 0.04$ dex. 

\begin{figure*} 
\begin{center}
\begin{tabular}{cc}
\includegraphics[scale=.75]{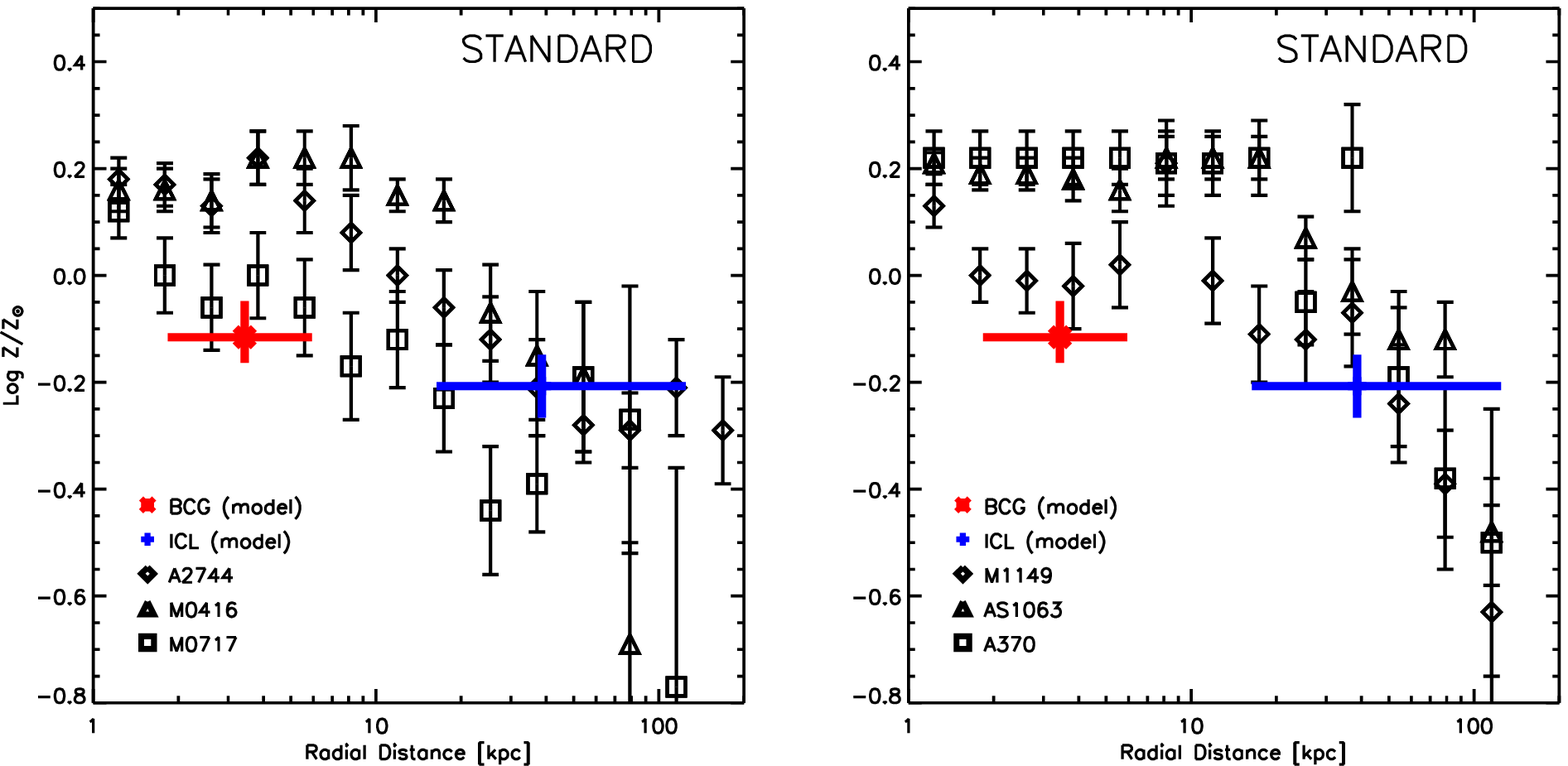} \\
\includegraphics[scale=.75]{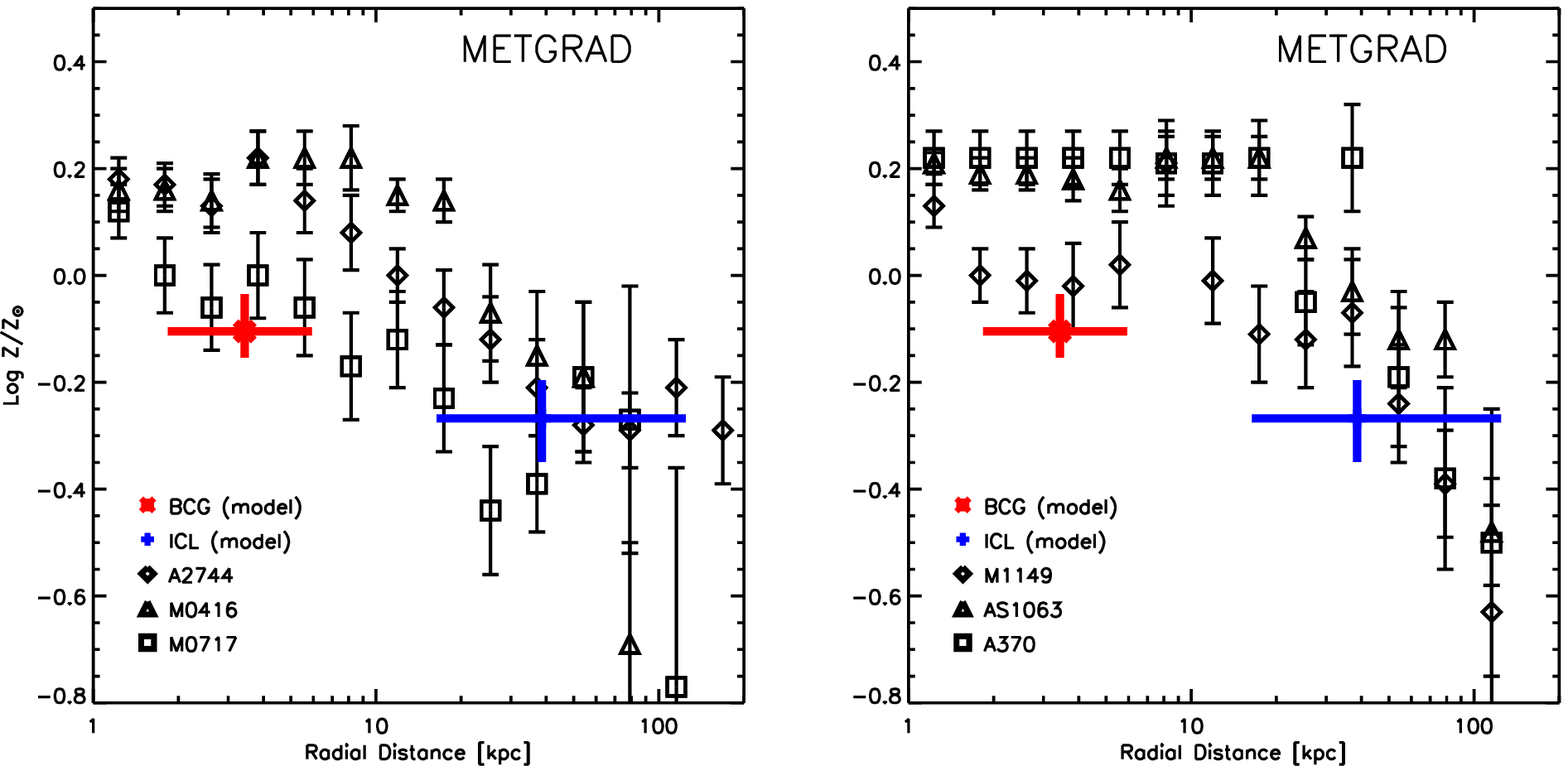} \\ 
\end{tabular}
\caption{Metallicity of BCGs and ICL as a function of radial distance for haloes at $z\sim 0.5$, compared to six observed profiles from \citealt{montes18} at the same redshift.
In order to make the plots more readable, we split the comparison with observed data in two panels for each model. The upper panels show the predictions of our STANDARD 
model, while the bottom panel show the predictions of model METGRAD (solid thick lines, which are the regions between 16$^{th}$ and 84$^{th}$ percentiles).}
\label{fig:gradient_distance}
\end{center}
\end{figure*}

In Figure \ref{fig:gradient_distance} we plot the radial metallicity profile of BCGs and ICL in six clusters in the redshift range $0.3<z<0.6$ (diamonds, triangles and squares in each panel) by \cite{montes18}, 
and compare them with predictions (solid thick lines which represent the regions between the 16$^{th}$ and 84$^{th}$ percentiles) of our two models (top panels for the {\small STANDARD} model and bottom panels 
for the {\small METGRAD} model). In semi-analytic models we do not have spatial information, which means that radial profiles are not available and must be assumed. To compare our results with observed data, 
we collect the metallicity of all BCGs (no trend with halo mass so we can increase the statistic by considering all of them rather than those in haloes of similar mass as those observed), and consider them as 
two-component systems: bulge and disk. In order to place them in the plot, we derived a mass-weighted radius of the galaxy (considering the bulge and disk radii) and concentrated the whole amount of metals in 
it. This avoids assumptions on the radial metallicity profile which would inevitably bias the results. We compute the mass-weighted radius of the BCG as follows:

\begin{equation}\label{eqn:eq_rbcg}
 R_{BCG} = \frac{R_{bulge} \cdot M_{bulge}^* +1.68\cdot R_{sl}\cdot M_{disk}^*}{M_{BCG}^*} \, ,
\end{equation}
where $R_{bulge}$ and $1.68\cdot R_{sl}$ are the half mass radii of the bulge and disk, $M_{bulge}^*$ and $M_{disk}^*$ are the masses of the bulge and disk, respectively.

The ICL is placed in the plot in a similar way (solid thick lines on the right). We assume that the ICL dominates at $10\cdot R_{sl}$ and place in the plot its metallicity at the distance 
equivalent to $10\cdot R_{sl}$. A caveat must be noted. Our predictions are integrated results that consider the metallicity of the whole system (BCG or ICL), while observed data refer to the metallicity at a given 
radial distance from the BCG center. Overall, both our model predictions agree reasonably well with observed data, and show little differences (as expected from Figure \ref{fig:bcgicl_met}). If we assume that 
the BCG+ICL systems predicted by our models have some kind of negative gradient, model data would move up in the inner regions and down in the region dominated by the ICL. Thus, our predictions have to be considered 
as a lower limit in the BCG dominated region, and as a upper limit in the ICL dominated region. Our model predictions agree well also with the observed data by \cite{demaio15}, who find typical metallicities ranging 
from $\sim$ [0.0,0.15] dex at 10 kpc, and $\sim$ [-0.4,-0.1] dex at 100 kpc (see their Fig.13), and with data by \cite{montes14}, who find similar metallicities as those just quoted (see their Fig.2).

Our models show similar results but, if compared to the observed data, the assumption of a radial metallicity gradient in satellites makes the model predictions to go to the right direction, i.e. higher metallicities 
for the BCGs and lower metallicities for the ICL. We will come back on this in Section \ref{sec:discussion}.

\subsection{ICL and BCG colors}
\label{sec:colors}

\begin{figure*} 
\begin{center}
\begin{tabular}{cc}
\includegraphics[scale=.65]{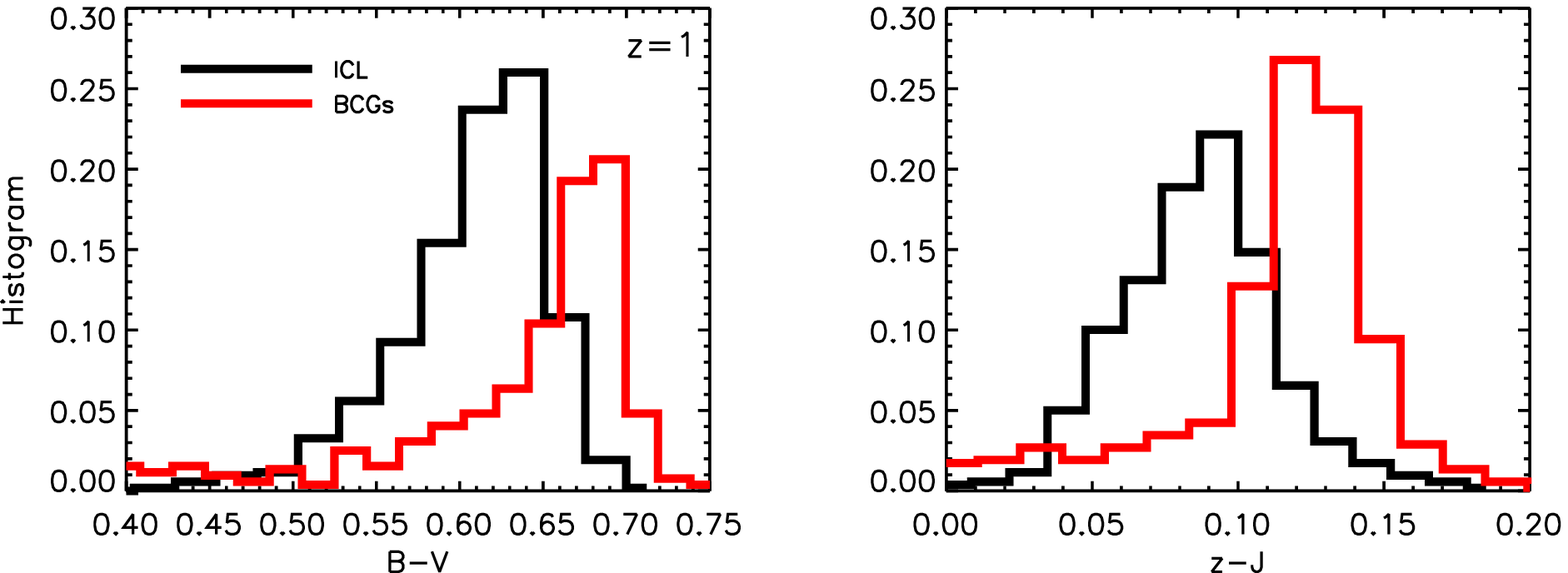} \\
\includegraphics[scale=.65]{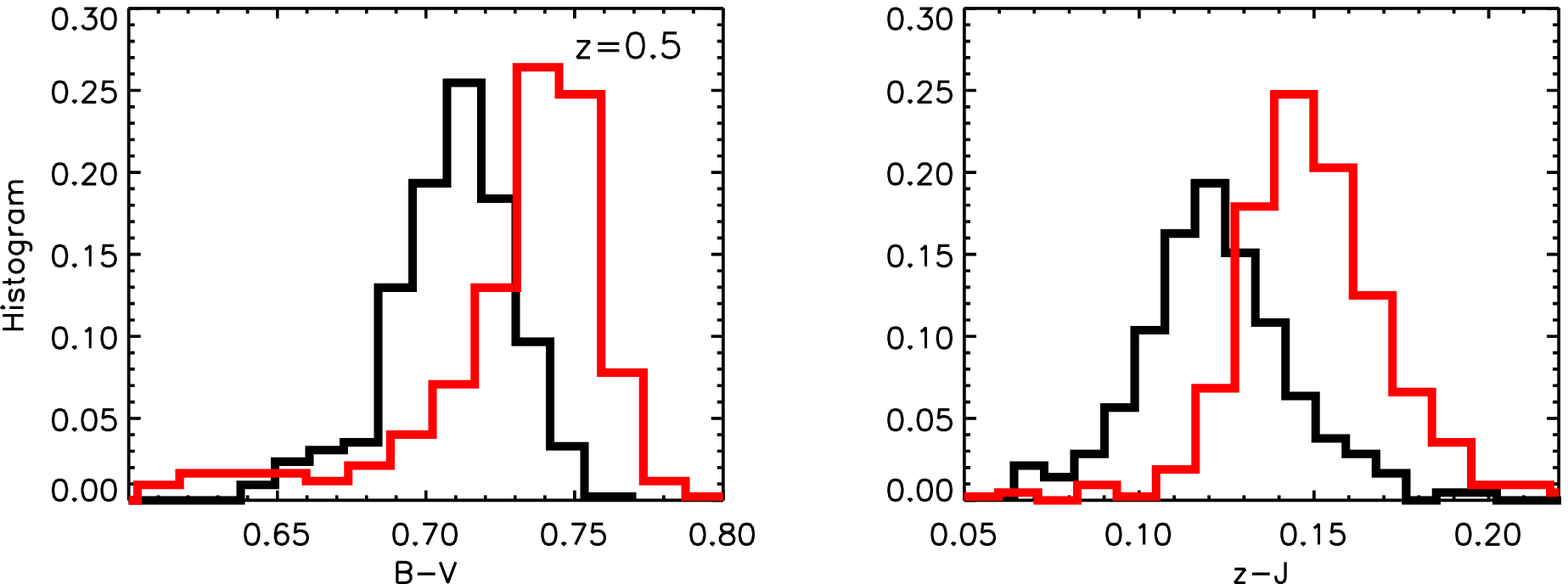} \\ 
\includegraphics[scale=.65]{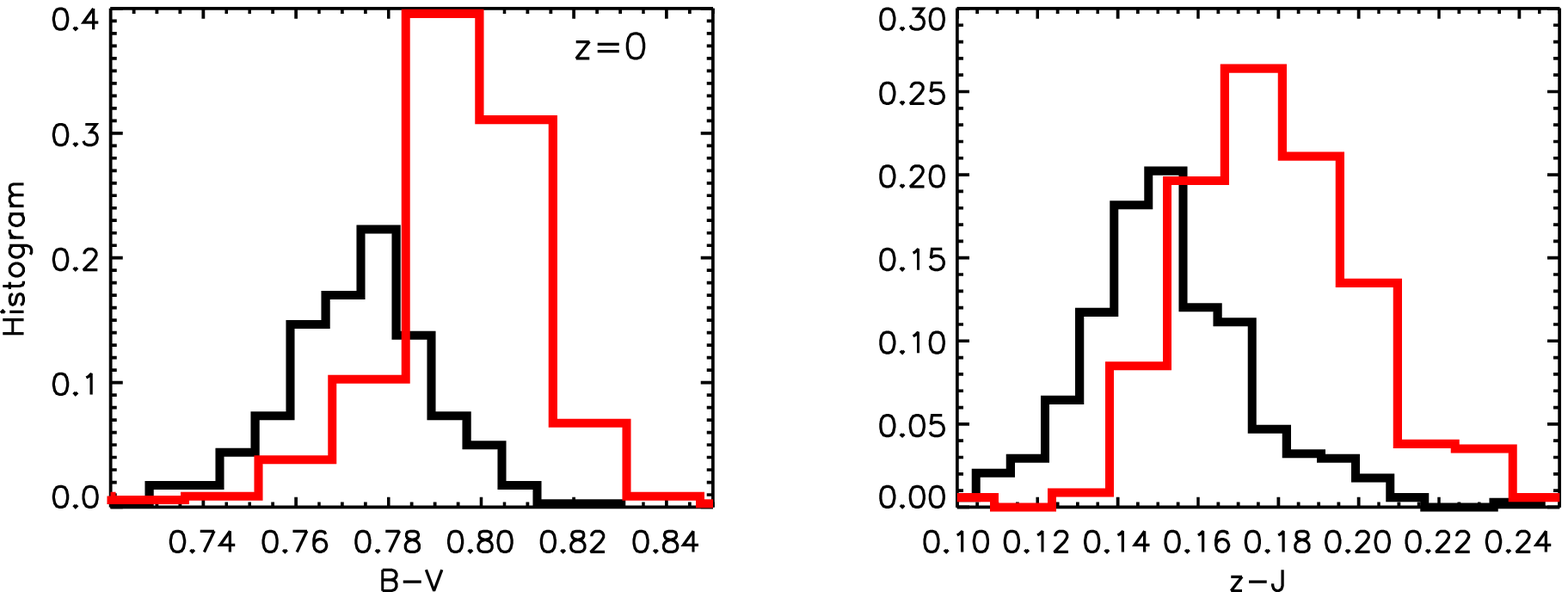} \\
\end{tabular}
\caption{B-V (left panels) and z-J (right panels) color distributions for BCGs (red histograms) and ICL (black histograms), at $z=1$ (upper panels), $z=0.5$ (middle panels),
and $z=0$ (bottom panels).}
\label{fig:histo_BVzJ}
\end{center}
\end{figure*}

In this section we analyse the ICL and BCG colors in the BVJgriz system. As seen in Section \ref{sec:metallicity}, the predictions of our models are very similar. In the analysis that follows we find negligible 
differences between the two models, then, for the sake of shortness, we show the results from our {\small STANDARD} model only.

\begin{figure*} 
\begin{center}
\begin{tabular}{cc}
\includegraphics[scale=.65]{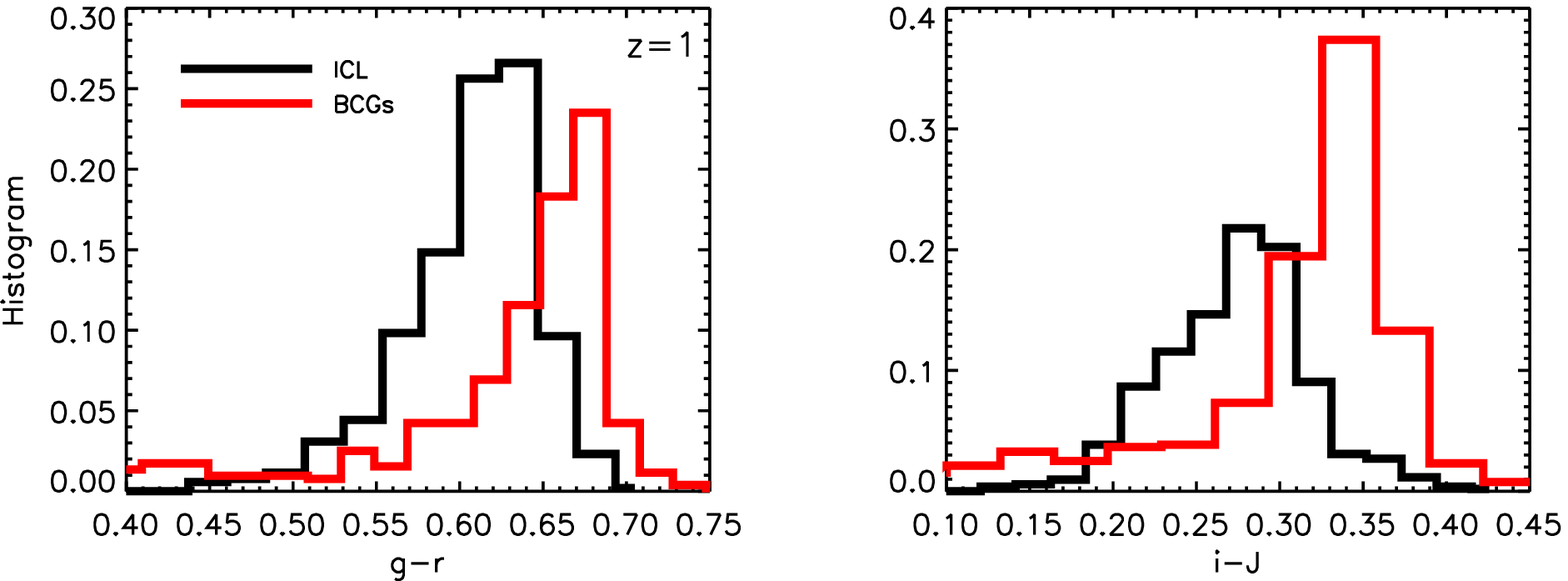} \\
\includegraphics[scale=.65]{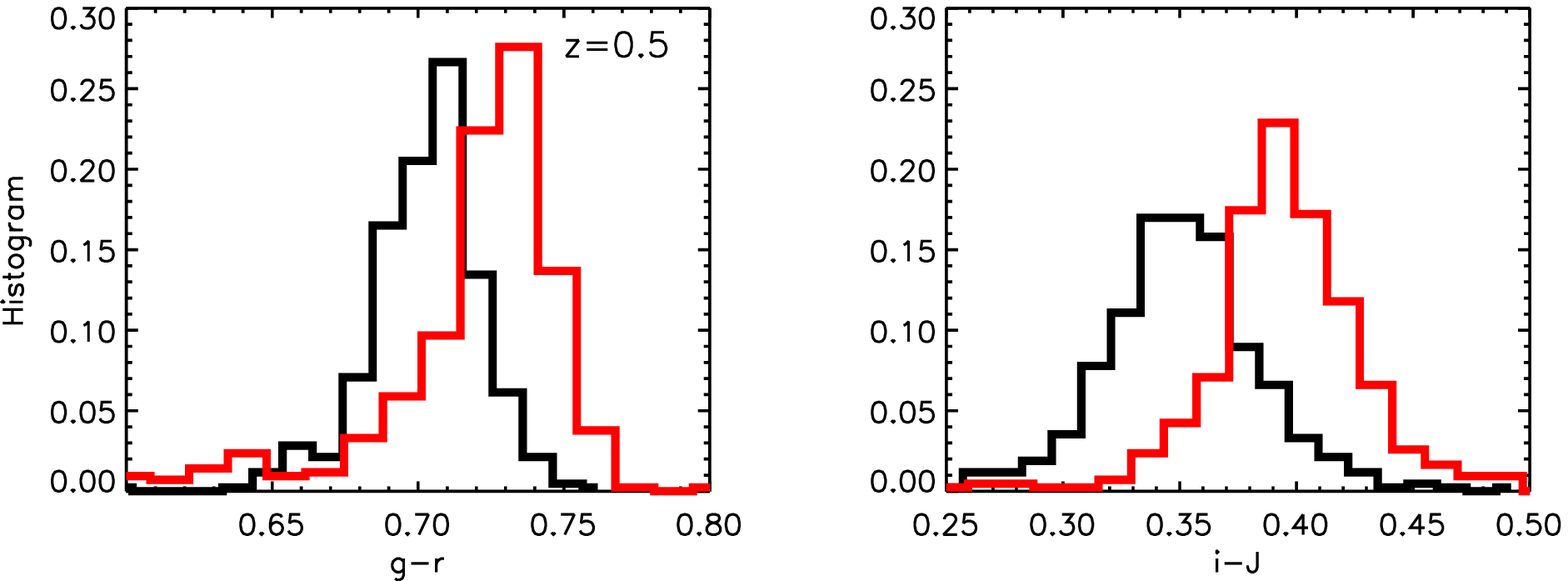} \\ 
\includegraphics[scale=.65]{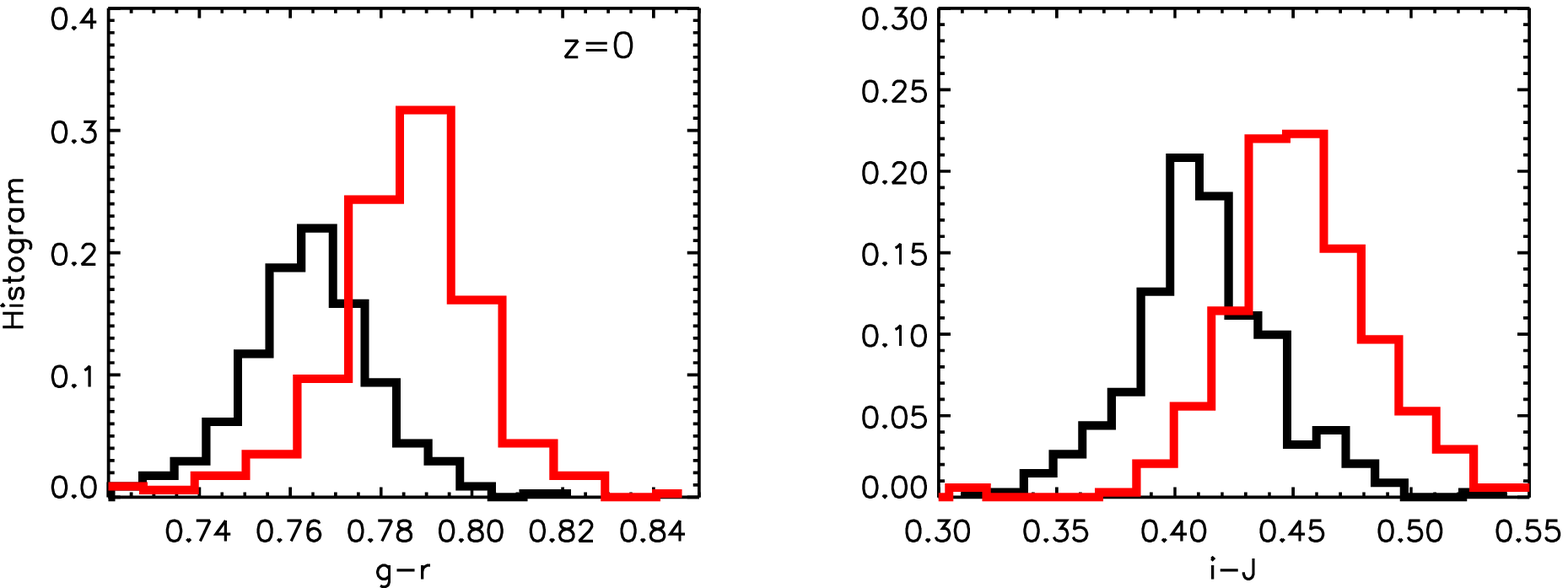} \\
\end{tabular}
\caption{Same as Figure \ref{fig:histo_BVzJ} for g-r color (left panels), and i-J color (right panels).}
\label{fig:histo_griJ}
\end{center}
\end{figure*}

In Figure \ref{fig:histo_BVzJ} we plot the histogram of B-V (left panels) and z-J (right panels) colors of BCGs (red lines) and ICL (black lines), at redshifts $z=1$ (top panels), $z=0.5$ (middle panels) and 
$z=0$ (bottom panels). Similarly, in Figure \ref{fig:histo_griJ} we plot g-r and i-J colors. The distributions of the four colors show that BCGs are redder than the ICL, at any redshift since the ICL starts to form 
(which we consider to be $z \sim 1$ as shown in C14 and C18). However, in all cases the color difference between BCGs and ICL is less than $\sim 0.1$ mag, translating in a mild color gradient which does not 
weaken with decreasing redshift. 

Our results are consistent with several observational data. \cite{demaio18} study the color gradient of 23 galaxy groups and cluster in the redshift range $0.3<z<0.9$ and find that the 
BCG+ICL color gets bluer towards the region dominated by the ICL, indicating that the ICL is bluer than the BCG. In their Fig.5 they plot the BCG+ICL color profiles ordered by increasing 
redshift. That plot shows mild gradients in most of the cases and no clear dependence on redshift, consistent with what we find in Figure \ref{fig:histo_BVzJ} and \ref{fig:histo_griJ}.

\begin{figure*} 
\begin{center}
\begin{tabular}{cc}
\includegraphics[scale=.4]{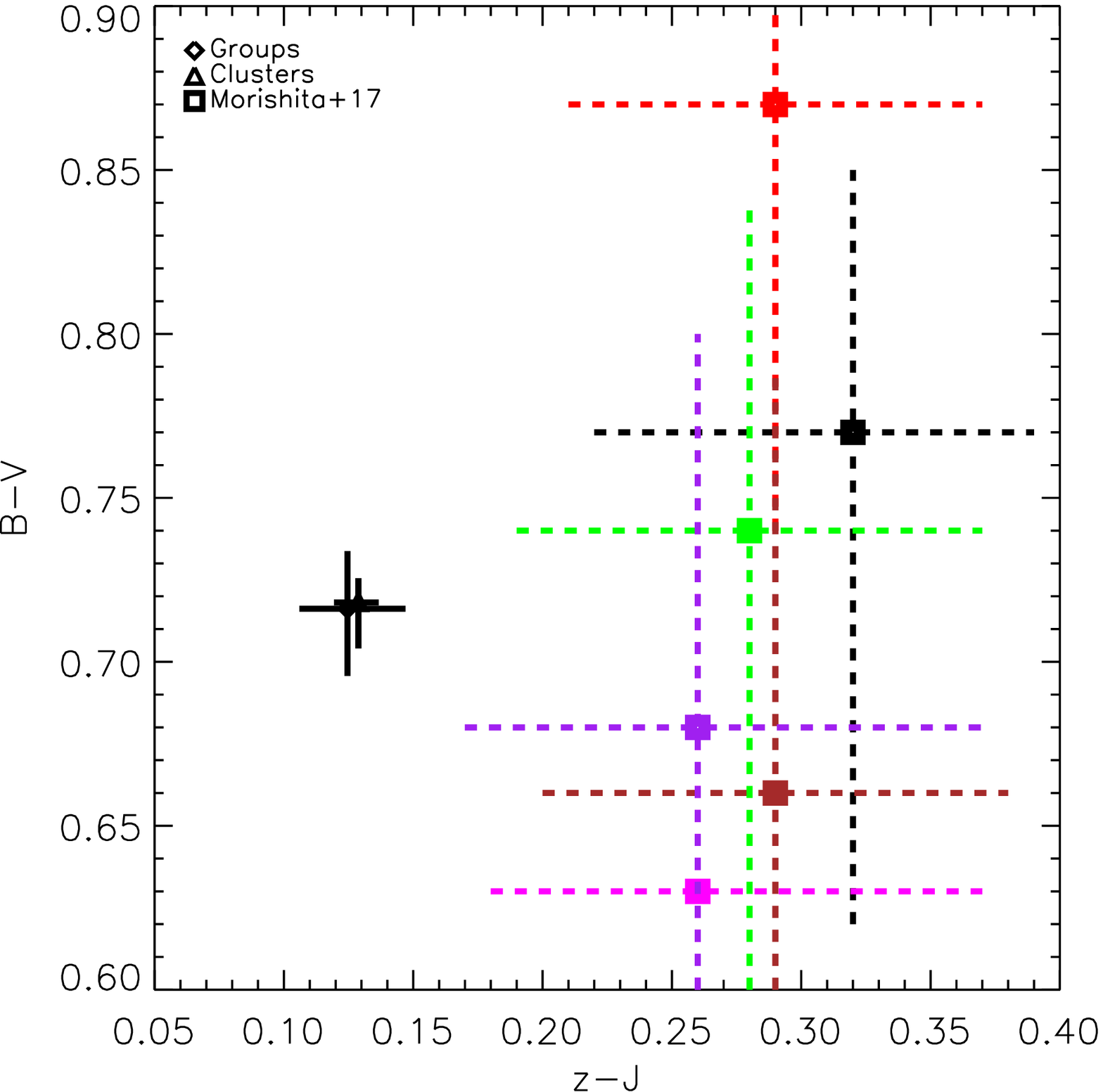} &
\includegraphics[scale=.4]{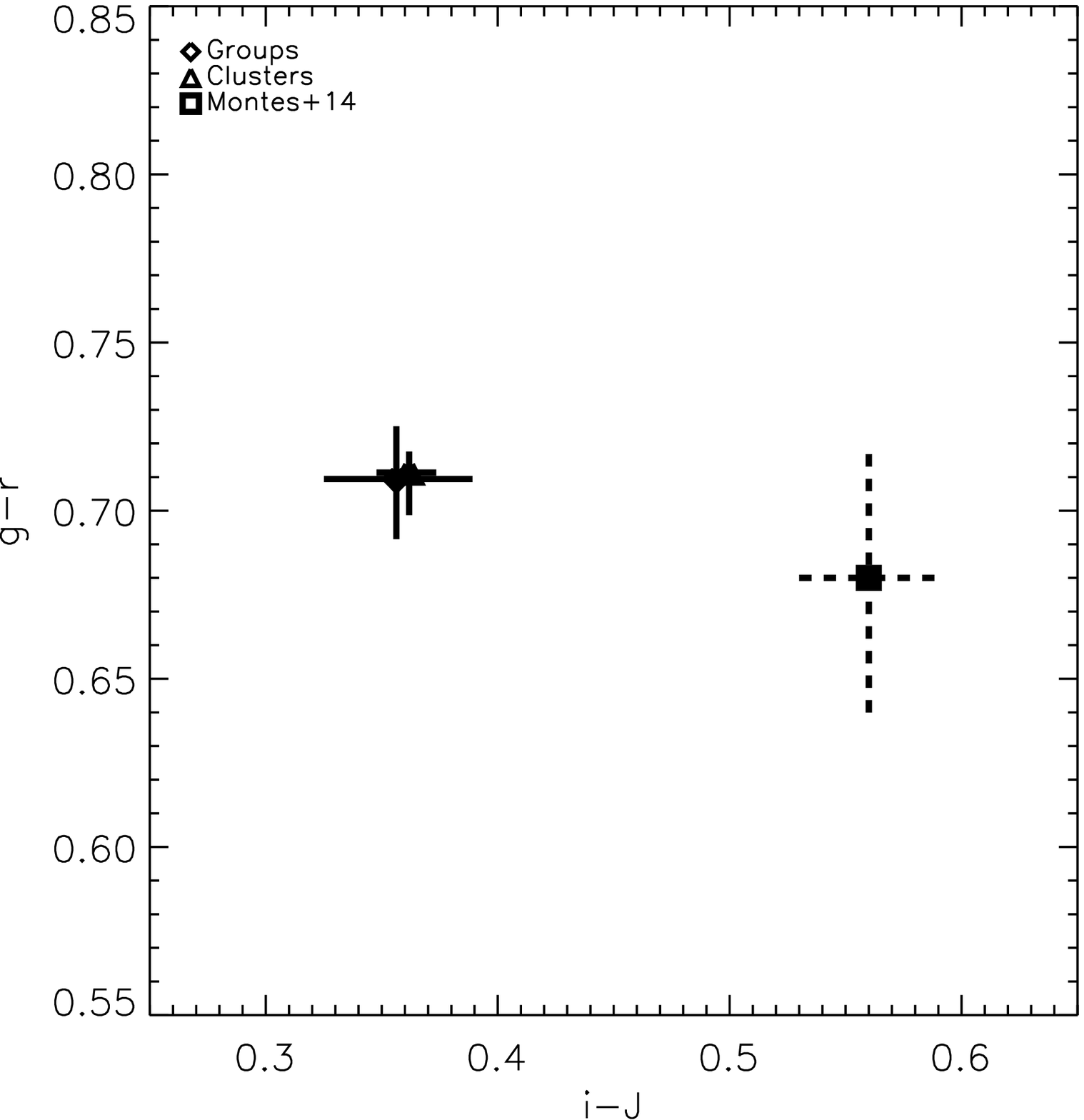} \\ 
\end{tabular}
\caption{Left panel: BVzJ color diagram of ICL in groups (diamond) and clusters (triangle) at redshift $z\sim0.5$, compared with observed data from \citealt{morishita17} (dashed lines which represent the region
between $\pm 1-\sigma$ scatter) at similar redshifts. Right panel: griJ color diagram of ICL in groups and clusters at redshift $z\sim 0.3$ (same simbols as in the left panel), compared with observed data from 
\citealt{montes14} (dashed lines which represent the region between $\pm 1-\sigma$ scatter). Solid thick lines represent the regions between 16$^{th}$ and 84$^{th}$ percentiles.}
\label{fig:coldiagram}
\end{center}
\end{figure*}

\cite{morishita17} investigate the ICL in six clusters at redshifts $0.3<z<0.6$ (the same clusters analysed by \citealt{montes18} and shown in Figure \ref{fig:gradient_distance}) and find clear negative color 
gradients (see their Fig.4). Qualitatively speaking, their results compare well with ours. Despite that we find similar B-V colors ($\sim 0.7$ mag), our z-J ICL colors are comparable with theirs only at large 
distances, around 300 kpc. As stated above, we do not have spatial information and, as in the case of the metallicity, radial color profiles are possible only making assumptions which would bias the results. If 
we focus the attention on the innermost regions, their ranges in B-V are consistent with ours, but again, our z-J colors are bluer. The comparison with observed data in g-r and i-J colors is very similar. 
Our g-r ICL colors are consistent with the color g-r$=0.68 \pm0.04$ mag of the Abell Cluster 2744 at $z=0.3$ (\citealt{montes14}), and with g-r$\sim 0.7$ mag of the Fornax cluster (\citealt{iodice17}), but our 
i-J colors are bluer by around 0.2 mag when compared with i-J$\sim 0.55$ of the Abell Cluster 2477 (\citealt{montes14}). 

\begin{figure*} 
\begin{center}
\begin{tabular}{cc}
\includegraphics[scale=.40]{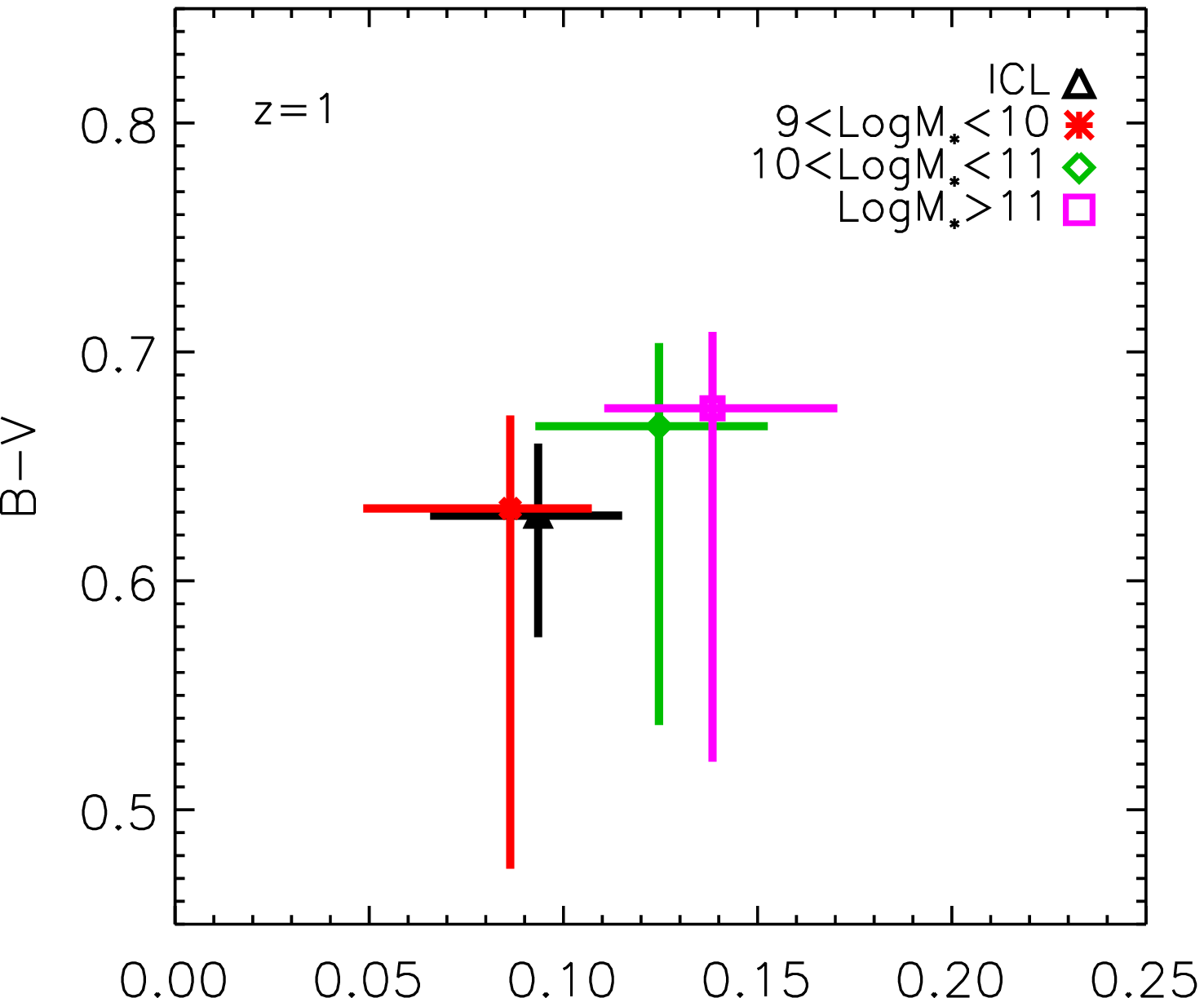} &
\includegraphics[scale=.40]{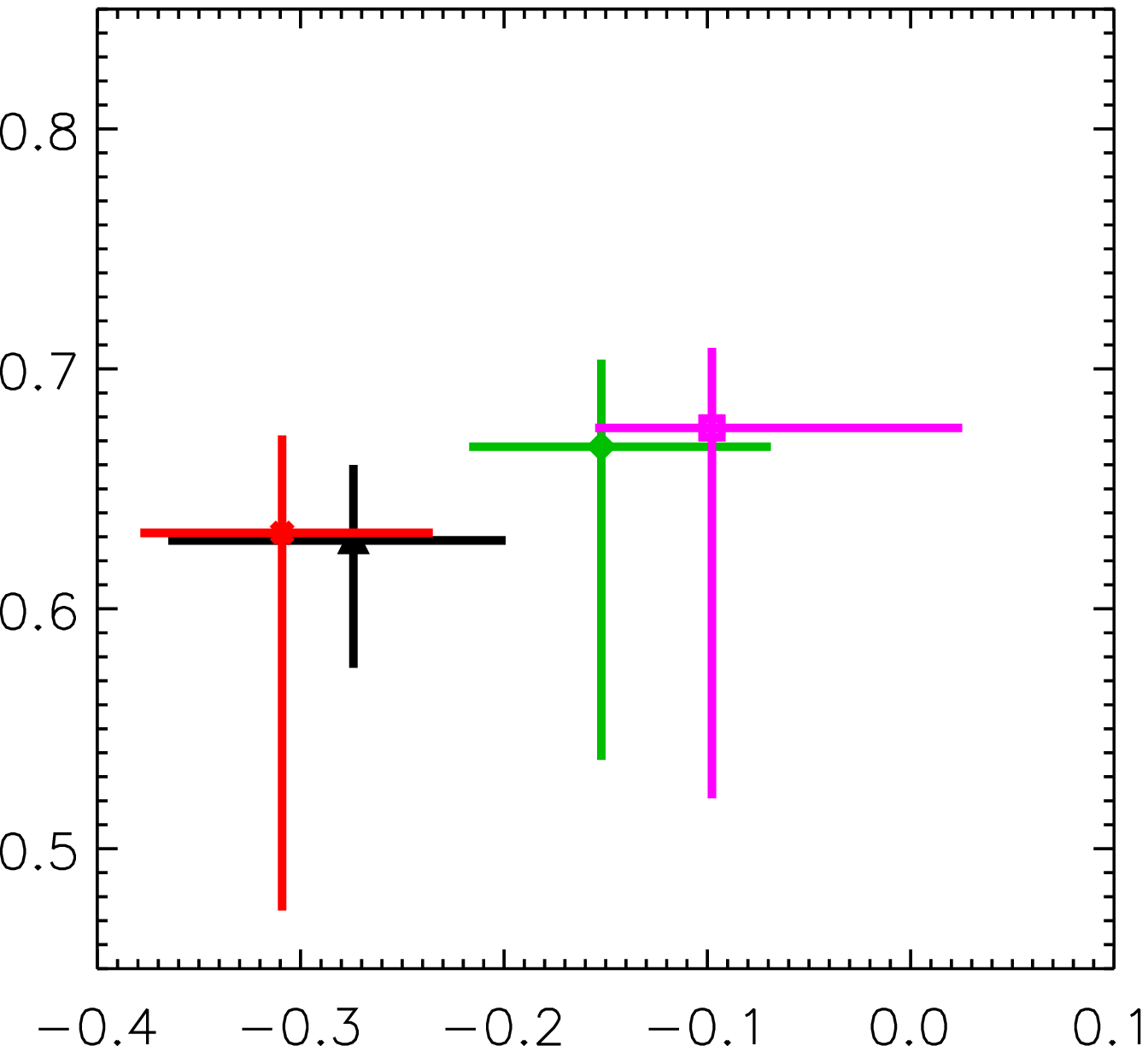} \\
\includegraphics[scale=.40]{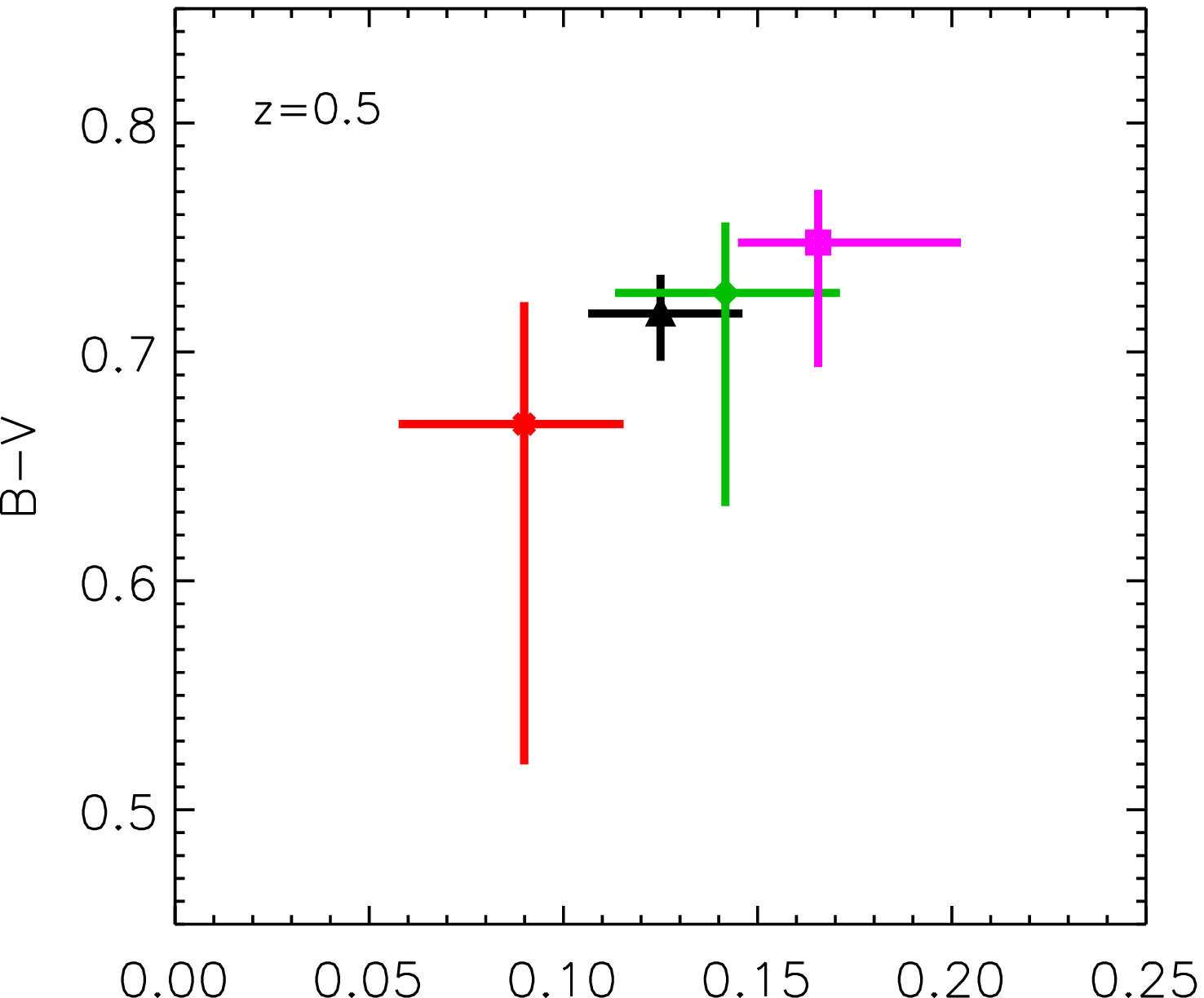} & 
\includegraphics[scale=.40]{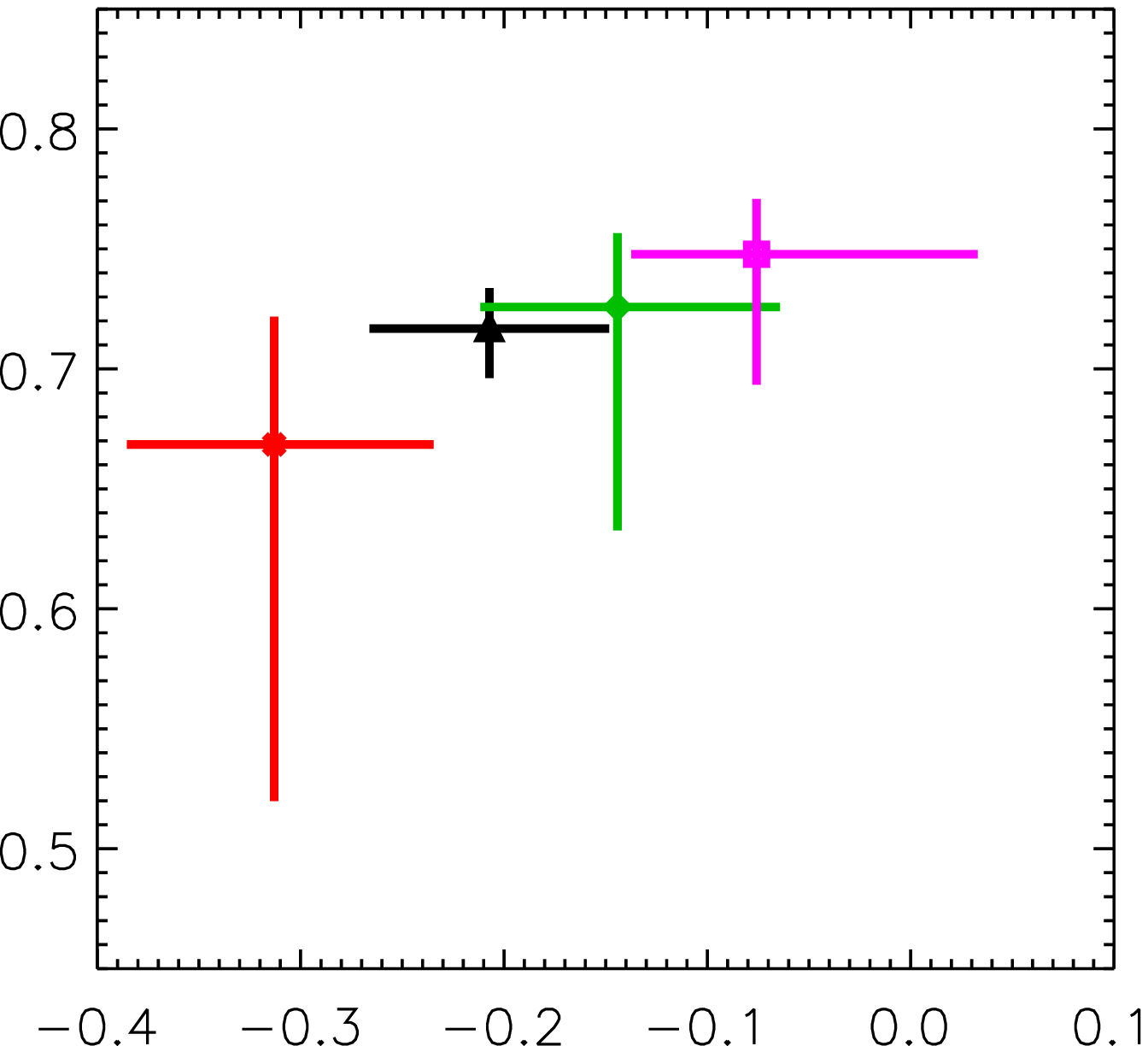} \\
\includegraphics[scale=.40]{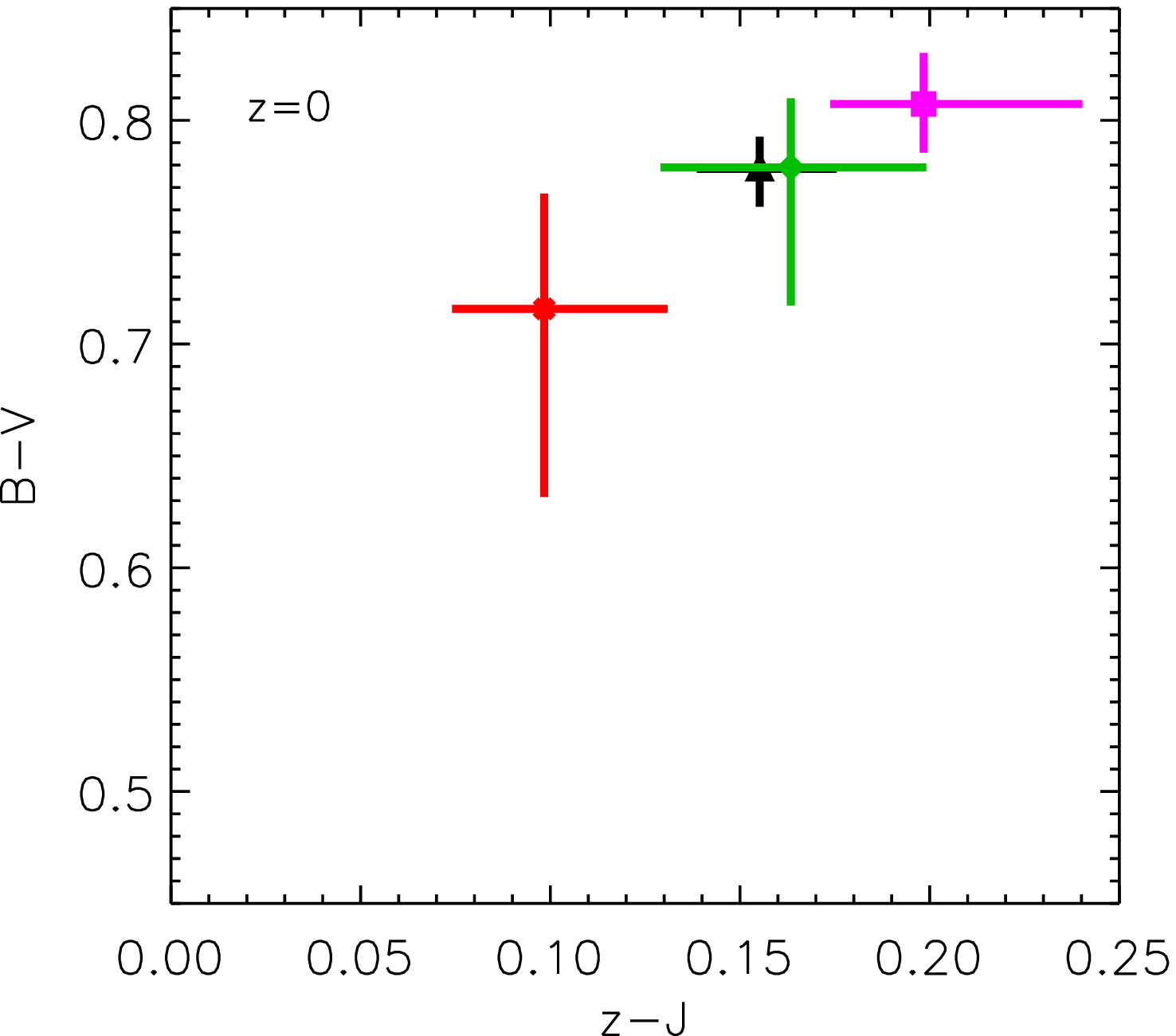} &
\includegraphics[scale=.40]{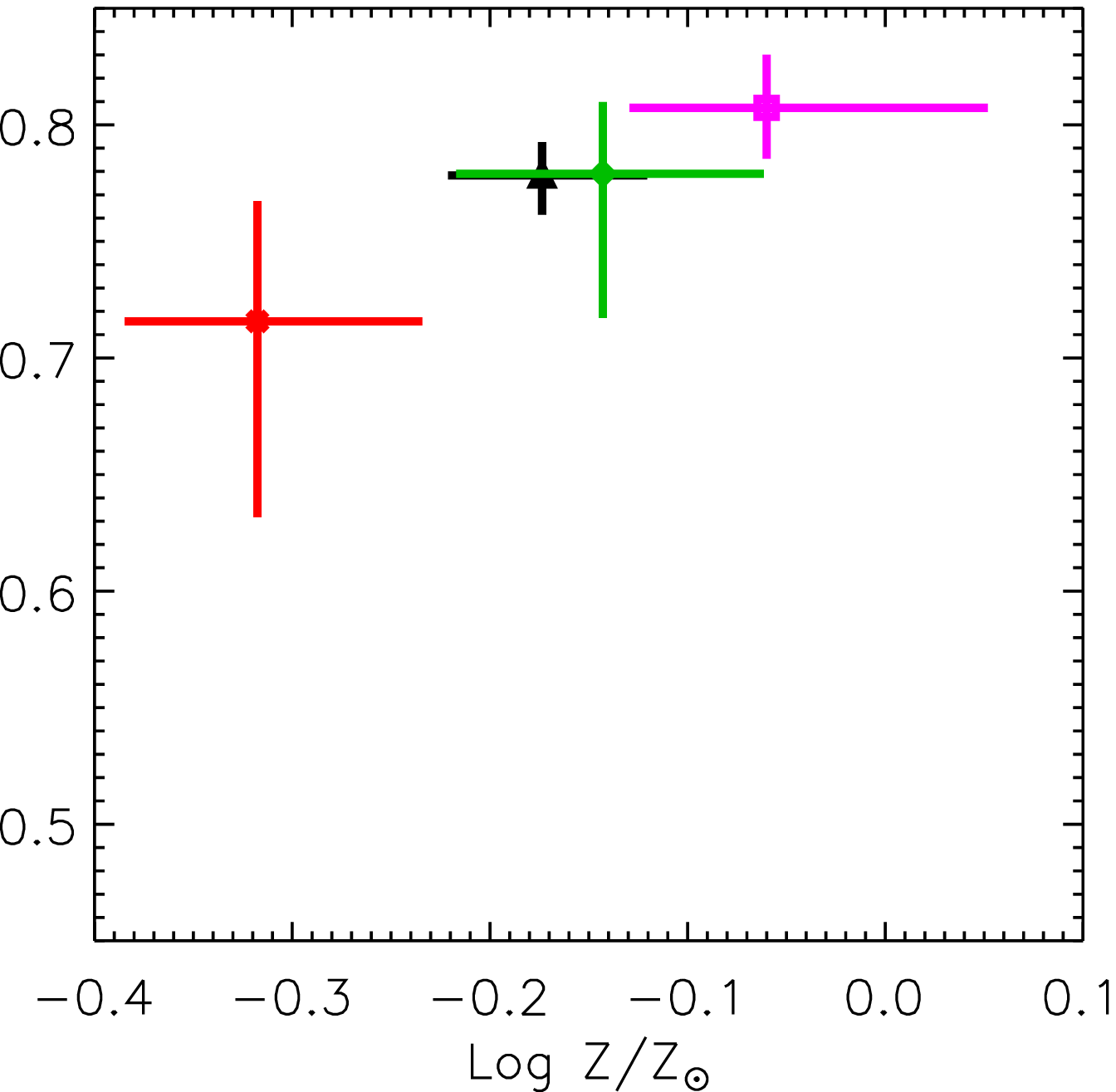} \\
\end{tabular}
\caption{Left panels: BVzJ color diagram at different redshifts (different panels) for the ICL (black triangle), and satellite galaxies in different stellar mass ranges: $9<\log M_* <10$ 
(red star), $10<\log M_* <11$ (green diamond), and $\log M_* >11$ (magenta square). Right panels: same as the left panels but the z-J color as been replaced by the metallicity. Solid thick lines represent 
the regions between 16$^{th}$ and 84$^{th}$ percentiles.}
\label{fig:colors_sat}
\end{center}
\end{figure*}
For a more simplistic representation and a more quantitative comparison of the results cited above, we show in Figure \ref{fig:coldiagram} the BVzJ (left panel) and griJ (right panel) color diagram, collecting 
the predictions of the {\small STANDARD} model for galaxy groups ($\log M_{halo} < 14.5$, diamonds), galaxy clusters ($\log M_{halo} \geq 14.5$, triangles), observed data by \cite{morishita17} (color lines and squares in the left 
panel), and data by \cite{montes14} (black lines and square in the right panel). Data by \cite{morishita17} cover a wide range both in B-V and z-J, while the observation by \cite{montes14} in g-r is narrower.
As stated above, our predictions agree fairly well with observed data in B-V and g-r, while there is a non-negligible offset between model and observations in z-J ($\sim 0.15$ mag), 
and i-J ($\sim 0.2$ mag). We will come back on this issue in Section \ref{sec:discussion}.

\subsection{Color-color and color-metallicity planes}
\label{sec:planes}

The color-color plane has been used in the past to understand from what kind of satellite galaxies the ICL acquires its mass, just by comparing the typical colors of the ICL with those of 
satellites in different ranges of stellar mass (\citealt{morishita17}). In the left panels of Figure \ref{fig:colors_sat} we plot the B-V and z-J colors of the ICL (triangles), and satellite galaxies in different 
mass ranges, $9<\log M_* <10$ (red stars), $10<\log M_* <11$ (green diamonds), and $\log M_* >11$ (magenta squares), as a function of redshift (different panels). The plots clearly show that satellites in the 
stellar mass range $9<\log M_* <10$ have colors similar to those of the ICL at $z=1$, suggesting that they are the systems which contribute most to the ICL in the beginning of its formation \footnote{Very recently,
\cite{ko18} find an observational hint for an earlier formation of the ICL in a galaxy cluster at redshift $z=1.24$. However, this cluster might be one of the exceptional cases and more statistic at these redshifts 
is needed.}. Quantitatively and qualitatively speaking, this result is in good agreement with the recent observation of the ICL F105W-F140W color of a galaxy cluster at $z\sim 1.2$ by 
\cite{ko18}, who find F105W-F140W $\sim 0.7$ mag. Although we do not show it, at the same redshift we find g-r$=0.62\pm0.03$ mag for the ICL, and g-r$=0.63\pm0.05$ mag for satellites in the stellar mass range $9<\log M_* <10$.

As time passes (see middle panel), the colors of the ICL get much closer to those of more massive galaxies, in the stellar mass range $10<\log M_* <11$. At the present time (bottom panel), intermediate/massive satellites are still the major contributors to the ICL.

The same conclusions can be drawn by looking at the color-metallicity plane (right panels). At $z=1$, both colors and the metallicities of low mass galaxies are very close to the color and metallicity of the ICL,
but, at lower redshifts and down to the present time, these ICL properties get closer and closer to those of intermediate/massive galaxies. Colors and metallicity then confirm the prediction maden in C14, i.e. 
intermediate/massive galaxies are responsible for the bulk of the stellar mass and metals in the ICL.

\cite{morishita17}, in a similar way, compared the B-V and z-J colors of the ICL in six clusters at an average redshift $z\sim 0.5$ with the typical colors of satellite galaxies in different stellar mass ranges. 
In tension with our predictions, they find that low mass, $\log M_* \lesssim 10$, galaxies are likely the most important source for the ICL, since its colors are more consistent with the colors of those galaxies. 
However, as noted by the authors themselves, the tension can be explained by the presence of strong color gradients in massive galaxies. We will fully discuss this important point in Section \ref{sec:discussion}.

\section{Discussion}
\label{sec:discussion}

The primary goal of this study is to focus on colors and metallicities of the ICL and BCGs and show how our model predictions compare with observational data. The standard version of the model adopted here has been 
developed in C14, where we have shown the basic properties of the ICL, including its metallicity at the present time. At that time no many observational measurements 
of colors and metallicities were available, and a combined (theory and observation) analysis focused on using observed data to constrain theoretical models was not possible, or completely reliable. Since then, an 
important amount of data has been collected, and not only at the present time where we focused our attention in the analysis done in C14, but at higher redshifts. Currently, observations have already reached $z \sim 1$
and promising campaigns are starting to go further, beyond $z=1$. Having a full coverage of observed properties of the ICL from $z \sim 1$ and down to the present time is strictly necessary for a solid comprehension of the 
formation and evolution of the ICL, and it is, at the same time, extremely helpful for setting theoretical models of galaxy formation. 

In Section \ref{sec:results} we have presented a series of results aimed to test our models against available observations. As stated in C14 and mentioned in this paper, colors and metallicities are important quantities that can 
tell us more about the ICL formation and its evolution, and then on the dynamical state and history of the cluster in which the ICL is found (\citealt{feldmeier04,krick06,murante07,purcell07,puchwein10,rudick11,contini14,contini18,burke15,groenewald17}). 
In the following we discuss in detail our results and their implications for the general picture of the ICL formation that the collection of different works is shaping. 

The metallicities of BCGs and ICL have been studied by several authors (e.g., \citealt{feldmeier98,durrell02,williams07,loubser09,coccato11,montes14,demaio15,montes18}). Overall, these studies find the stellar ages of the ICL
being between $2-13$ Gyr and sub-solar ICL metallicities in the range $-0.8<\log Z/Z_{\odot}<-0.2$, and hints of the presence of radial metallicity gradient in the BCG+ICL system. An important question is: what can the radial 
metallicity profile tell us? The answer to this question has been discussed several times and by several authors (e.g. \citealt{montes14,demaio15,morishita17,demaio18,montes18,contini18} and others) and the key 
point relies on the main mechanism responsible for the formation of the ICL. 

According to the main literature on the topic, a handful of processes have been invoked, but currently only two are considered to be important sources of ICL: galaxy mergers and stellar stripping. The relative importance 
of each of them in contributing to the ICL stellar mass as a function of time has different consequences on the ICL properties, such as the metallicity (as well as colors). In fact, if we assume that mergers between satellites 
and the BCG are the main channel, we would not expect a clear metallicity (or color) gradient in the BCG+ICL system, simply because major or multiple minor mergers would mix the stellar populations and thus flattening the pre-existing gradient. On the 
other hand, if we assume that stellar stripping is the most important channel, we do expect some gradient, from super-solar metallicities in the BCG, to sub-solar metallicities in the ICL (i.e. a negative gradient). 
This is because stellar stripping removes stars from the outskirts of the satellites, which are more metal-poor than the average system, and the typical values strongly depend on what kind of galaxies (in terms of stellar mass) 
contribute most. BCGs lies on the right side of the mass-metallicity relation, and so they are, on average, more metal-rich than satellites. \emph{Ergo}, if ICL stars come from stripping of satellites that are more metal-poor, 
and, on the top of it we add the fact that stripping acts on the outskirts of the satellites, the net result would be an ICL more metal-poor than BCGs. 

These arguments have been used by several authors in the last few years. \cite{montes14} (but see also \citealt{montes18}), based on the colors of the Abell Cluster 2744, derive a mean metallicity of the ICL slightly sub-solar. 
According to the properties of the stellar population in the ICL, they conclude that most of it formed via disruption of galaxies with mass and metallicity comparable to those of the Milky Way. Similarly, \cite{demaio15} use 
stellar population synthesis models to convert the observed colors to metallicity of four clusters at $z\sim 0.5$. They find negative metallicity gradients from super-solar (BCGs) to sub-solar (ICL), which they explain as the 
result of tidal stripping of $L^*$ galaxies and thus ruling out major mergers as the main contributors. In a later study, \cite{demaio18}, with a more numerous sample of galaxy clusters (23) in a wider range of redshift ($0.3<z<0.9$), 
strengthen their previous conclusion by ruling out the contribution to the ICL from dwarf galaxies as the major channel. In fact, as discussed also in C18, in order to reproduce the observed luminosity of the ICL, the number of disruption 
events of this kind of galaxies would considerably flatten the faint-end slope of the luminosity/stellar mass functions after $z<1$ (when the ICL starts forming), at odds with observations 
(e.g. \citealt{mancone12,ilbert13,muzzin13,tomczak14} and others). 

Observational clues in favor of the major merger scenario are also present in the recent literature, albeit strengthened by theoretical arguments. \cite{burke15} focus on the BCG stellar mass growth from $z \sim 0.9$ to $z \sim 0.1$
and assume that, at each merger between the BCG and the satellite, 50\% of the stellar mass of the satellite galaxies goes to the ICL, thus finding a BCG and ICL grow factors in line with the expectations from theoretical models
(C14, \citealt{murante07}). Similarly, \cite{groenewald17} address the same point, between $0.1\lesssim z\lesssim 0.5$. These authors make the same assumption for the percentage of mass that moves to the ICL 
(50\%) and conclude that major mergers can explain the growth rate of BCGs, and at the same time they bring enough stellar mass to the ICL down to the present day. As noted in C18, they make use of the stellar mass growths 
published in C14, which consider both stellar stripping and mergers. Although their method is inconsistent, their find a similar growth factor. 

Our models favor the stellar stripping channel rather than mergers (C14, C18), and their predictions are in line with the picture described above. One key point of this work relies on the importance of a 
metallicity gradient in satellite galaxies (our {\small METGRAD} model). As shown in Section \ref{sec:results}, assuming a metallicity gradient in satellites that contribute to the stellar mass and metals in the ICL has just a 
little effect on the ICL and BCG metallicities (see Figure \ref{fig:bcgicl_met} and \ref{fig:gradient_distance}). Nevertheless, this assumption brings the predictions towards the right direction, that is, BCGs more metal-rich and
ICL more metal-poor. However, a caveat is worth noting. As argued in C14, the mass-metallicity relation predicted by our models is offset low with respect to the observed one (e.g., \citealt{gallazzi05}) at the massive end. In 
particular, the observed BCG metallicities, which are similar to those of the most massive galaxies, are expected to be at least 0.2-0.3 dex higher (e.g., \citealt{vonderlinden07}). Modelling the ICL does not substantially 
improve the disagreement in the massive end, despite it goes to the right direction (higher metallicities for more massive galaxies).

For the first time in semi-analytics we present predictions of the ICL colors. As discussed in Section \ref{sec:colors}, B-V and g-r colors are in good agreement with the observed ones, and we find BCGs to be slightly redder 
than the ICL, as observed. Nevertheless, our z-J and i-J colors are bluer compared with observations. In both colors we find an offset of around 0.2 mag, which probably depends on the  
response of the J filter, considering that in all the other bands our results agree fairly well with observational data. However, despite the offset (which propagates from the stars in galaxies to the stars in the ICL), our models predict mild radial color gradients in the BCG+ICL system at any redshift since $z\sim 1$, 
in agreement with observations (e.g., \citealt{montes14,demaio15,iodice17,demaio18}).

Colors are a useful tool to understand the channels that contribute most to the ICL. In C14 we show that most of the ICL comes from intermediate/massive galaxies and its metallicity is very similar to that of these galaxies. 
In that study we focus our attention to the present time, when all the ICL is formed. Here, as shown in Figure \ref{fig:colors_sat}, we present the same information as a function of redshift by comparing the colors/metallicity 
of the ICL with those of satellite galaxies in different ranges of stellar mass, similarly to \cite{morishita17}. In the beginning of its formation the ICL is mainly built-up by relative low mass galaxies ($9<\log M_* <10$), but 
already at $z\sim 0.5$ and down to the present time, more massive ($10<\log M_* <11$) satellites play the most important role. If we compare our results at $z\sim 0.5$ with the observed data by \cite{morishita17} at similar redshifts, 
we find a disagreement. In fact, \cite{morishita17} show that low mass satellites ($\log M_* <10$) have colors closer to those of the ICL, and conclude that these are the main contributors to the ICL. However, the presence of color 
gradients in massive satellites might reconcile (at least partly) our disagreement, because stripping acts mainly in the outskirts of satellites, regions typically bluer than the integrated colors (see, e.g., \citealt{morishita15}). 
Then, although they find ICL colors closer to those of relative low mass satellites, the main contribution can come from stars stripped from the outskirts of massive satellites. As discussed in C14, dynamical friction arguments 
fully support this picture, being more massive satellites faster in reaching the innermost regions of the halo and more likely to be subject to stellar stripping than low mass galaxies.

\section{Conclusions}
\label{sec:conclusions}

We have coupled a semi-analytic model of galaxy formation with a set of N-body simulations to make predictions of colors and metallicities of BCGs and ICL. In the analysis we took advantage of the prescription for the 
formation of the ICL presented in C14 and C18, and a modification of it which considers a metallicity gradient in satellite galaxies that are subject to stellar stripping. We compared the results presented in Section \ref{sec:results}
and discussed in Section \ref{sec:discussion} to test our model against current theories for the formation and evolution of the ICL. In the light of our results and their implications, we conclude the following:
\begin{itemize}
 \item BCGs are more metal-rich than the ICL. Moreover, the assumption of a metallicity gradient in satellite galaxies subject to stellar stripping brings prediction to the right direction (BCGs more metal-rich and ICL 
       more metal-poor), but does not have a significant impact on the results, quantitatively speaking (Figure \ref{fig:bcgicl_met}).
 \item Both prescriptions predict a negative metallicity gradient and mild color gradients of the BCG+ICL system, in good agreement with observed data (Figure \ref{fig:gradient_distance}, \ref{fig:histo_BVzJ} and \ref{fig:histo_griJ}). 
       B-V and g-r ICL colors are well reproduced, but our z-J and i-J ICL colors are offset-low with respect observations by $\sim 0.2$ mag (Figure \ref{fig:coldiagram}). 
 \item The contribution to the ICL in terms of stellar mass and metals come from galaxies of different mass, depending on the redshift. At the beginning of its formation, the ICL acquires most of the mass from galaxies in the 
       stellar mass range $9<\log M_* <10$, but already at $z\sim 0.5$ and down to the present time intermediate/massive galaxies with mass in the range $10<\log M_* <11$ contribute most (Figure \ref{fig:colors_sat}).
\end{itemize}

Future observational campaigns designed to measure colors and metallicity of BCG+ICL in a wide range of redshift can put some constraints on the formation and evolution of the ICL. In C18 we argue that a possible solution to 
the debate stellar stripping/mergers as the main channel can be found on cluster scales in the local Universe, by separating the ICL associated to the BCG from that formed and somehow linked to satellite galaxies. For the reasons 
discussed above, we conclude this study highlighting the importance of radial colors and metallicity gradients in support of stellar stripping as the main channel. The first measurements are suggesting this picture, but we need 
more data to finally confirm or prove it wrong.

\section*{Acknowledgements}
S.K.Y. and E.C. acknowledge support from the Korean National Research Foundation 
(NRF-2017R1A2A1A05001116) and from the Brain Korea 21 Plus Program (21A20131500002). This study was 
performed under the umbrella of the joint collaboration between Yonsei University Observatory 
and the Korean Astronomy and Space Science Institute. X.K. acknowledges financial support by
the 973 Program (2015CB857003) and the NSFC (No.11333008).

\label{lastpage}

\end{document}